\documentclass[aps,prl,reprint,10pt,superscriptaddress]{revtex4-1} 
\pdfoutput=1
\usepackage[T1]{fontenc}
\usepackage[utf8]{inputenc}

\usepackage{graphicx}  
\usepackage{dcolumn}   
\usepackage{bm}        
\usepackage{amssymb}   
\usepackage{amsmath}
\usepackage{url}
\usepackage{natbib}
\usepackage{gensymb}

\usepackage[normalem]{ulem}
\usepackage{soul}

\usepackage{graphicx}
\usepackage[dvipsnames]{xcolor}
\xdefinecolor{RED}{named}{BrickRed}

\newcommand{\p}{\partial}

\newcommand{\mbf}{\boldsymbol}

\newcommand{\bu}{\mbf{u}}
\newcommand{\bk}{\mbf{k}}
\newcommand{\bx}{\mbf{x}}

\usepackage{color} 

\definecolor{linkcolor}{rgb}{0,0,0.6} 

\begin{document}
\title{Inertial wave turbulence driven by elliptical instability}

\author{Thomas Le Reun}
\affiliation{Aix Marseille Univ, CNRS, Centrale Marseille, IRPHE UMR 7342, Marseille, France}
\author{ Benjamin Favier}
\affiliation{Aix Marseille Univ, CNRS, Centrale Marseille, IRPHE UMR 7342, Marseille, France}
\author{Adrian J. Barker}
\affiliation{Department of Applied Mathematics, School of Mathematics, University of Leeds, Leeds, LS2 9JT, UK}
\author{Michael Le Bars}
\affiliation{Aix Marseille Univ, CNRS, Centrale Marseille, IRPHE UMR 7342, Marseille, France}

\begin{abstract}

The combination of elliptical deformation of streamlines and vorticity can lead to the destabilisation of any rotating flow via the elliptical instability. 
Such a mechanism has been invoked as a possible source of turbulence in planetary cores subject to tidal deformations.
The saturation of the elliptical instability has been shown to generate turbulence composed of non-linearly interacting waves and strong columnar vortices
with varying respective amplitudes, depending on the control parameters and geometry. 
In this paper, we present a suite of numerical simulations to investigate the saturation and the transition from vortex-dominated to wave-dominated regimes.
This is achieved by simulating the growth and saturation of the elliptical instability in an idealised triply periodic domain, adding a frictional damping to the geostrophic component only, to mimic its interaction with boundaries.
We reproduce several experimental observations within one idealised local model and complement them by reaching more extreme flow parameters. 
In particular, a wave-dominated regime that exhibits many signatures of inertial wave turbulence is characterised for the first time.
This regime is expected in planetary interiors.

\end{abstract}

\maketitle

\date{\today}

The elliptical instability is a fundamental mechanism of importance in a wide range of fluid phenomena. 
Originally described in the context of strained vortices \cite{bayly_three-dimensional_1986}, it has been studied in various situations such as in vortex dipoles and in wakes. 
More generally, it has been proposed as a general process to transfer energy to smaller scales in turbulence
(see \cite{kerswell_elliptical_2002} and references within). 
It is important for geophysical fluid dynamics because it can drive flows in planetary cores subjected to tidal deformations \cite{lacaze_elliptical_2005,le_bars_coriolis_2007,le_bars_tidal_2010,le_bars_flows_2015}.
It has been invoked to explain the magnetic field of the early Moon \cite{le_bars_impact-driven_2011} and 
the Earth \cite{andrault_deep_2016}.   
This instability develops in rotating fluids when the streamlines are deformed into ellipses.
Its linear growth, due to the resonance of two inertial waves with the elliptical basic flow is well described both theoretically \cite{craik1989,le_dizes_three-dimensional_2000,kerswell_elliptical_2002,cebron_libration-driven_2014} and experimentally \cite{lacaze_elliptical_2005,le_bars_coriolis_2007,le_bars_tidal_2010,Eloy2000}.
The non-linear saturation of the elliptical instability remains poorly understood but it is the relevant regime to describe vortex core breakdown \cite{SCHAEFFER2010} as well as dissipation and magnetic field generation in planetary cores \cite{Kerswell1998}.
Simulations and experiments of the elliptical instability exhibit a variety of non-linear behaviours depending on the values of the control parameters (the ellipticity of the streamlines and the viscosity) 
and on the geometry.
The saturation of the instability  can lead to either sustained flows \cite{barker_mag_2014,grannan_experimental_2014,favier_generation_2015} or cyclic behaviours between laminar and turbulent states \cite{Malkus1989,Eloy2000,barker_non-linear_2013,Barker2016}, reminiscent of the ``resonant collapse'' of inertial waves observed by McEwan \cite{McEwan1970}.
The presence or absence of geostrophic modes appears to be important, but the diversity remains to be explained in detail.
Indeed, each inertial wave excited by elliptical instability of the base flow can be itself unstable to a triadic resonance with another pair of inertial waves \cite{kerswell_secondary_1999}: these secondary instabilities have been observed both numerically \cite{mason_nonlinear_1999,favier_generation_2015} and experimentally \cite{Eloy2003}. 
Whether these multiple resonances asymptotically lead to a 
wave turbulence regime \cite{galtier_weak_2003,bellet_wave_2006}, similar to the recently observed regimes with flexural waves in plates \cite{miquel2011}, gravity-capillary waves \cite{aubourg2015} and internal waves \cite{brouzet_energy_2016}, remains to be seen.
Barker and Lithwick \cite{barker_non-linear_2013}, in their local model of the elliptical instability, nonetheless showed that strong geostrophic flows emerge during the saturation and disrupt the inertial wave resonances, leading instead to growth and decay cycles.
This competition between dominant geostrophic modes and inertial waves is reminiscent of the duality observed in rotating turbulence where, on one hand, geostrophic flows are widely observed \cite{godeferd_structure_2015} and, on the other hand, inertial waves have been shown to survive on top of the turbulent background \cite{yarom_experimental_2014,clark_di_leoni_quantification_2014,campagne_disentangling_2015,favier_space_2010}.
Conversely to experiments and numerical simulations of rotating turbulence, where energy is injected arbitrarily into both vortices and inertial waves through an external artificial forcing, the elliptical instability provides a natural mechanism that initially injects energy into a few inertial waves only, whose properties can be theoretically predicted.

\begin{figure*}
    \centering
    \includegraphics[width=0.25\linewidth]{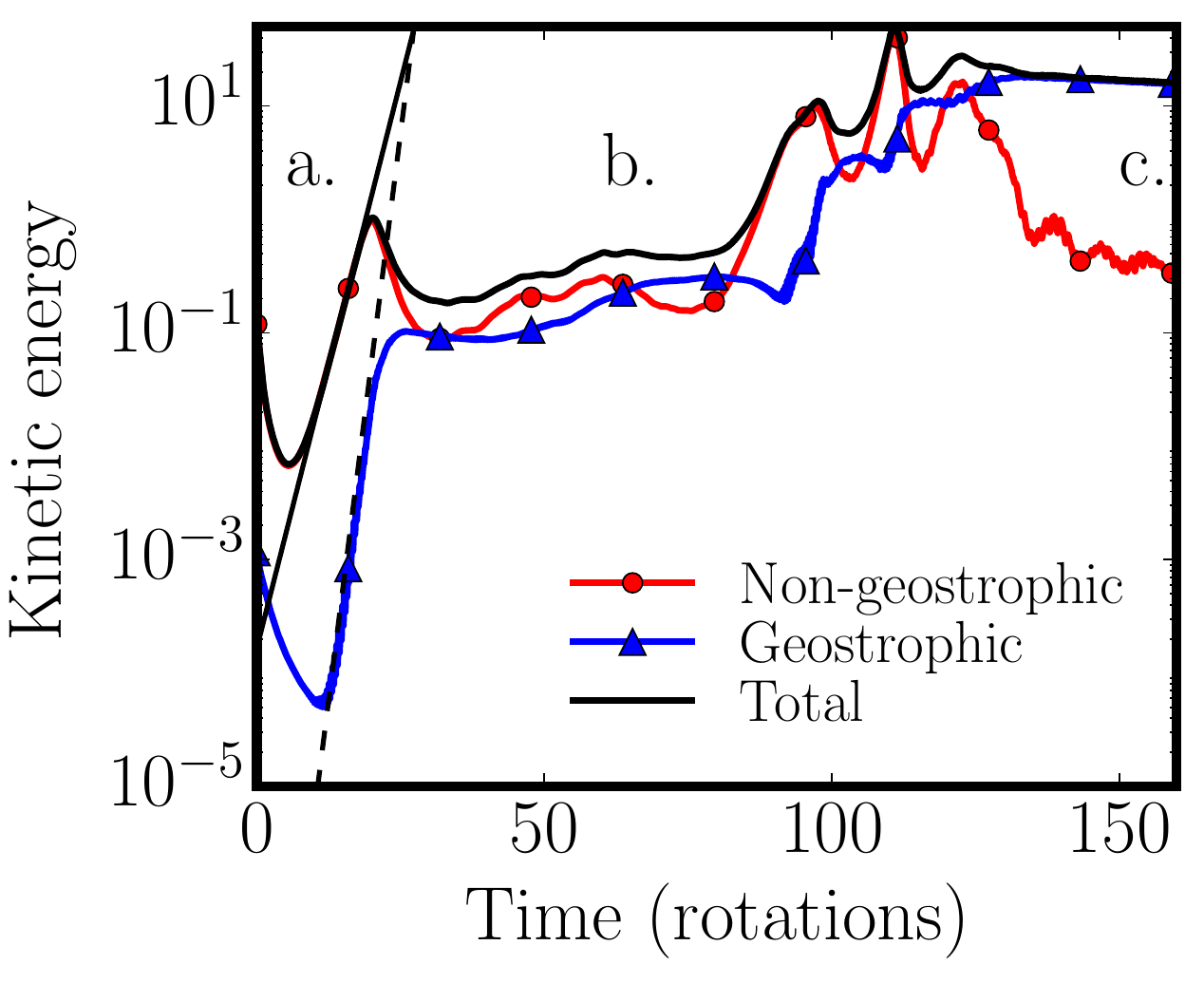}
    \includegraphics[width = 0.68\linewidth]{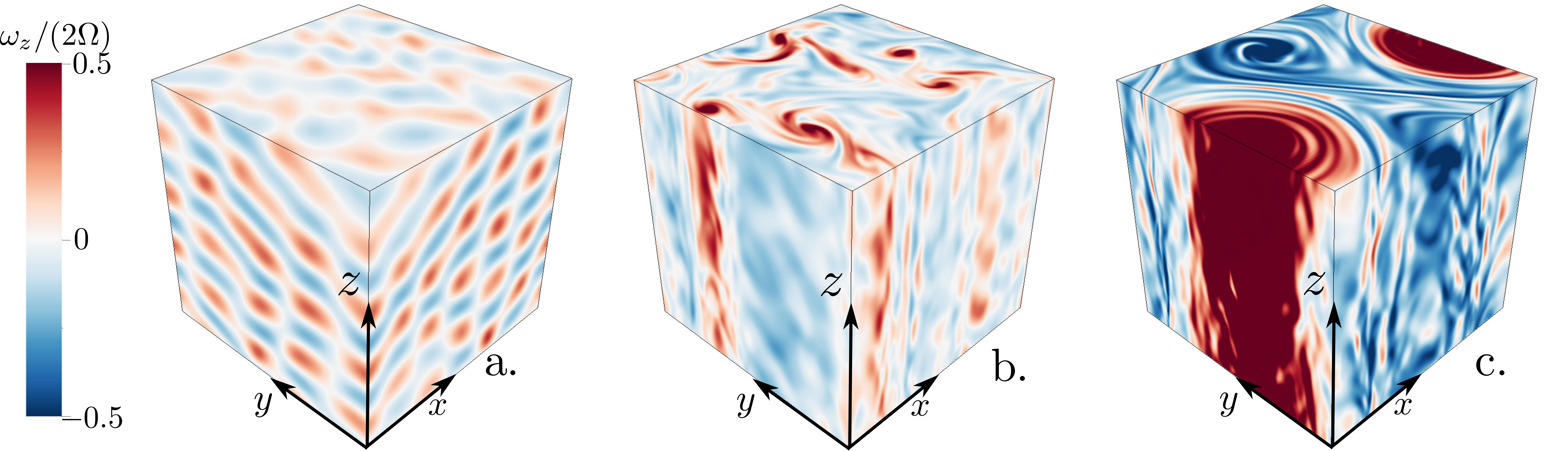}
    \vspace{-4mm}
    \caption{
    \textbf{Left:} typical evolution of the volume-averaged kinetic energy for $E= 10^{-5}$ and a $256^3$ resolution, from the exponential growth of a few waves (a.) to its non-linear saturation (b. and c.). Continuous and dashed lines account for exponential growth with rates $2 \sigma$ and $4 \sigma$ respectively with $\sigma$ the theoretical viscous growth rate of the instability. The velocity amplitudes are normalised by $k_{\rm{res}}^{-1} \Omega$.
    \textbf{Right:} corresponding typical snapshots of the vertical vorticity normalised by the background vorticity.
     }
\label{raw_results}
\end{figure*}

We focus here on the instabilities growing from a base flow made of solid body rotation at frequency $\Omega$ plus an elliptic deformation rotating at frequency $n$.
Both solid-body rotation and deformation are aligned with the vertical axis.
In the frame rotating with the deformation, the base flow  $ \mbf{U}_b $ reads  \cite{barker_non-linear_2013}:
\begin{equation}\label{eq:forcing}
    \mbf{U}_b  =  - \gamma \beta
    \begin{bmatrix}
    \sin(2\gamma t) & \cos(2\gamma t) & 0 \\
    \cos(2\gamma t) & -\sin(2\gamma t) & 0 \\
     0 & 0 & 0
    \end{bmatrix}
    \begin{bmatrix}
    x \\
    y\\
    z\\
    \end{bmatrix} = \mbf{A}(t) \mbf{x} 
\end{equation}
where we have introduced $\gamma ~\equiv ~ (\Omega- n)$ and $\beta$ the ellipticity.
This base flow can be seen as a local model of a periodically strained vortex \cite{le_dizes_three-dimensional_2000} or of a tidally deformed planetary core \cite{barker_non-linear_2013} and is a non-linear inviscid solution of the Navier-Stokes equations written in the rotating frame.
It is responsible for the parametric sub-harmonic excitation of two inertial waves with frequencies close to $\gamma$ provided $\vert\gamma\vert < 2 \Omega$ \cite{le_bars_flows_2015}. 
The dynamics of the perturbation $\bu$ around the base flow $\mbf{U}_b$ is governed by the following set of equations:
\begin{align}
    \label{eq:perturb_OI}
    \p_t \mbf{u} + (\mbf{U_b}.\mbf{\nabla}) \mbf{u} + (\mbf{u}\cdot\mbf{\nabla}) \mbf{U_b} +& (\mbf{u} \cdot \mbf{\nabla}) \mbf{u}  + 2 \mbf{\Omega} \times \mbf{u} \\
    &=- \mbf{\nabla} \Pi  + \nu \mbf{\nabla}^2 \bu & \nonumber\\    
    \label{eq:incomp}
    \mbf{\nabla}\cdot \mbf{u}~ &= ~0 
\end{align}
where $\Pi$ is the modified pressure, ensuring the incompressibility of the dynamics, and $\nu$ is the constant kinematic viscosity. 
We assume that the flow is homogeneous, 
thus enabling us to carry out pseudo-spectral direct numerical simulations of
equations (\ref{eq:perturb_OI})-(\ref{eq:incomp}) in a so-called periodic shearing box. 
This is achieved using the \textsc{Snoopy} code introduced by Lesur \cite{lesur_impact_2007} and adapted to the study of the elliptical instability by Barker \cite{barker_non-linear_2013}.
The perturbed flow is solved in a cubic box of size $L$ with periodic boundary conditions in all three directions. 
Lengths and time are normalised by $L$ and $\Omega^{-1}$ respectively. 
Simulations are initiated from a broad-band noise for wave numbers $4 \le k/(2\pi) \le 20$, though we obtain the same results if we instead adopt white noise.
The  control parameters are the normalised differential rotation $\gamma/\Omega$, the ellipticity $\beta$, which can be regarded as an input Rossby number, and the local Ekman number based on the box size $E \equiv \nu/(\Omega L^2)$. 
In the following, we choose $\beta = 5 \times 10^{-2}$, $10^{-6} \le E \le 10^{-5}$ and $\gamma/\Omega$ is set to $1.5$. 
The spatial resolution is up to $512$ grid points in each direction (see details in Supplementary Materials).  
Contrary to previous global DNS \cite{favier_generation_2015,Cebron2010}, we focus on values of $\beta$ as small as possible \textcolor{black}{as it is below $10^{-3}$ in planetary interiors 
}. 
The value of $E$ is a compromise between our desire to make the flow fully turbulent but have the simulation well resolved.
This number is usually between $10^{-15}$ and $10^{-10}$ in planetary cores essentially because of their massive size (see Supplementary Materials, which includes Refs \cite{Cebron2010,cebron_elliptical_2012}, for further information).
The ratio $\gamma/\Omega$ is set sufficiently high to avoid time-scale separation between the forcing and rotation.
The growth rate of the elliptical instability is an increasing function of $\gamma\beta$ \cite{kerswell_elliptical_2002}; the value 1.5 produces rapid turbulent saturation and ensures the selection of a mode with reasonably small wavelength compared to $L$.
Other forcing frequencies have been considered (see \textit{e.g.} Supplementary Materials for $\gamma/\Omega=1$) and the results are qualitatively unchanged.

\begin{figure*}
    \centering
    \includegraphics[width=0.68\linewidth]{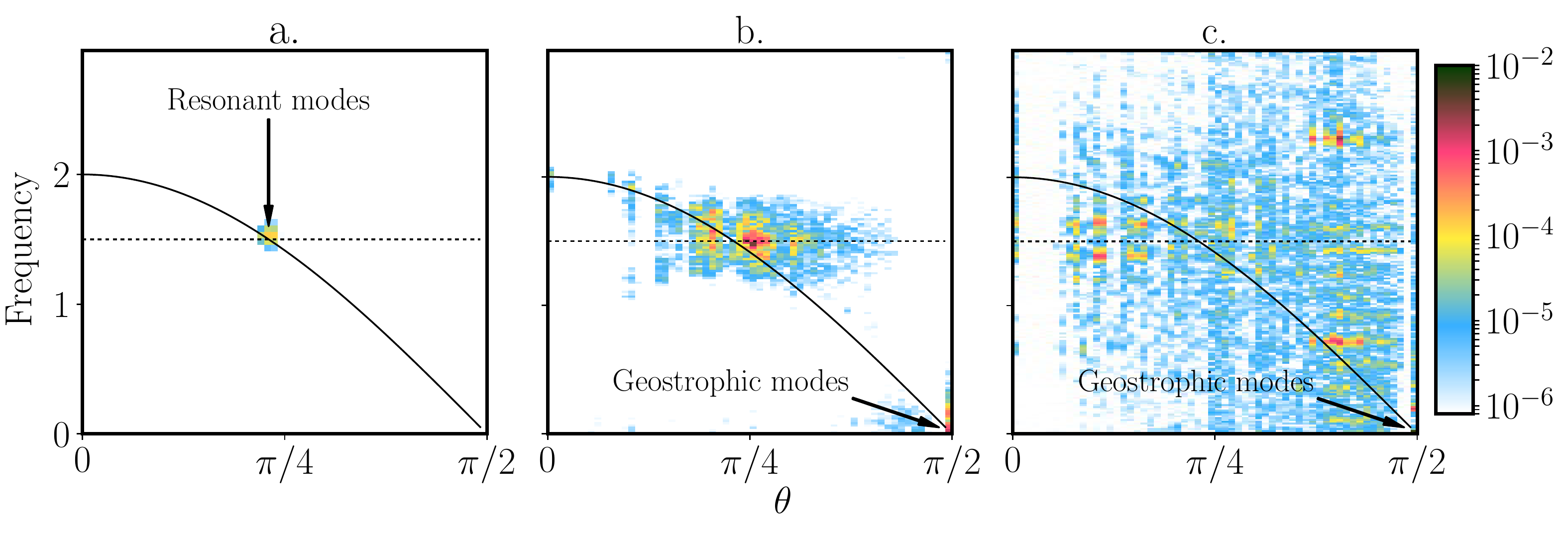}
    \vspace{-5mm}
    \caption{Spectral energy (normalised by $(k_{\rm{res}}^{-1} \Omega)^2$) as a function of the frequency in spin units and the angle between the wave vector and the rotation axis $\theta$ for $k/(2\pi) \in [3,20]$. The three figures correspond to the three stages identified in Figure \ref{raw_results}.
    The continuous line represents the inertial wave dispersion relation and the dotted one highlights the resonant frequency $\gamma/\Omega = 1.5$.
    }
    \label{fig:fr0_reldisp}
\end{figure*}
Snapshots of the evolving flow are displayed in Figure \ref{raw_results}, with the kinetic energy of the fluctuating velocity field, decomposed into the geostrophic 
and the residual non-geostrophic 3D components.
As previously shown in \cite{barker_non-linear_2013}, we first observe the exponential growth of a few planar waves (Figure \ref{raw_results}.a) whose wave number $k_{\rm{res}}/(2\pi )$ is between $6$ and $7$, thus ensuring a reasonable scale separation between the small-scale resonating waves and the size of the periodic box.
The growth rate of the instability is consistent with theoretical estimates \cite{craik1989,le_dizes_three-dimensional_2000} taking into account the bulk viscous damping.
The geostrophic component grows at twice the instability growth rate, which confirms that it is driven by the direct nonlinear interactions of the resonant inertial waves, and not by a secondary instability \cite{kerswell_secondary_1999}.
While the nonlinear interactions of inertial waves cannot lead to geostrophic modes with asymptotically low Rossby \cite{Greenspan1969}, this is not true for our simulation at finite Rossby number. 
In the saturated phase, Figure \ref{raw_results}.b  reveals the emergence of strong columnar vortices aligned with, and invariant along the axis of rotation, as observed in \cite[Fig. 5]{barker_non-linear_2013}.
The energy stored in the geostrophic modes is comparable to the energy in the rest of the flow. 
Similarly to what is observed in forced rotating turbulence \cite{Campagne2014,yarom_experimental_2014} or rapidly-rotating thermal convection \cite{stellmach2014,favier2014,guervilly2015}, a non-local inverse cascade of the geostrophic modes takes place until, at $t \simeq 120$ rotations onwards, one single large-scale vortex, or condensate, remains (Figure \ref{raw_results}.c).
Consequently, the kinetic energy of 
the non-geostrophic component drops by two orders of magnitude.

Although they are the primary structures during the exponential growth phase of the instability, the latter result shows that inertial waves are no longer the main contribution to the flow during the late stages of the saturation. 
Following \cite{yarom_experimental_2014,clark_di_leoni_quantification_2014}, we analyse how the kinetic energy is located around the dispersion relation of inertial waves given by
%
\textcolor{black}{
 $   \omega = \pm 2 \cos \theta \,$
%
}
where $\omega$ is the dimensionless frequency and $\theta$ is the angle between the wave vector and the rotation axis. 
This is achieved by computing the spatio-temporal Fourier transform $\hat{\bu}(\bk,\omega)$ of the velocity field $\bu(\bx,t)$ and summing contributions associated with the same angle $\theta$.
The spectral energy $|\hat{\bu}(\theta,\omega)|^2$ resulting from this approach is displayed in Figure \ref{fig:fr0_reldisp}. 
During the growth phase  ($t=10$ to $t= 30$ in Figure \ref{raw_results}), most of the energy is localised on the dispersion relation at the frequencies $\pm\gamma/\Omega$ (Figure \ref{fig:fr0_reldisp} a.), as expected from the temporal resonance condition with the base flow frequency $2\gamma$ \cite{le_bars_flows_2015,kerswell_elliptical_2002}.
In the early times of the saturated phase  ($t= 50$ to $t= 105$ in Figure \ref{raw_results}), the geostrophic flow grows in amplitude and departure from the dispersion relation is observed while the energy remains localised close to frequencies around $\gamma/\Omega$ (Figure \ref{fig:fr0_reldisp}b).
We interpret this result as follows: the base flow excites one particular frequency (fixed by the initial resonance condition with the base flow) while the wave vectors adapt themselves to interact with the growing geostrophic modes, whose energy is localised at $\theta\approx\pi/2$.
The geostrophic flow is dominantly cyclonic, as commonly observed in rotating turbulence \cite{godeferd1999,smith_near_2005}, and the increased apparent vorticity induces a shift toward larger values of $\theta$.
As the geostrophic vortices grow in amplitude, in phase c. (from $t=140$ onwards in Figure \ref{raw_results}) no inertial wave can be clearly identified (Figure \ref{fig:fr0_reldisp}c). 
Correspondingly, the non-geostrophic kinetic energy decreases, as observed in Figure \ref{raw_results}. 
After a viscous timescale, the geostrophic correction to the base flow is eventually dissipated leading to another resonance, and so on (more details about these cycles can be found in \cite{barker_non-linear_2013}).
\begin{figure*}
    \includegraphics[width=0.49\linewidth]{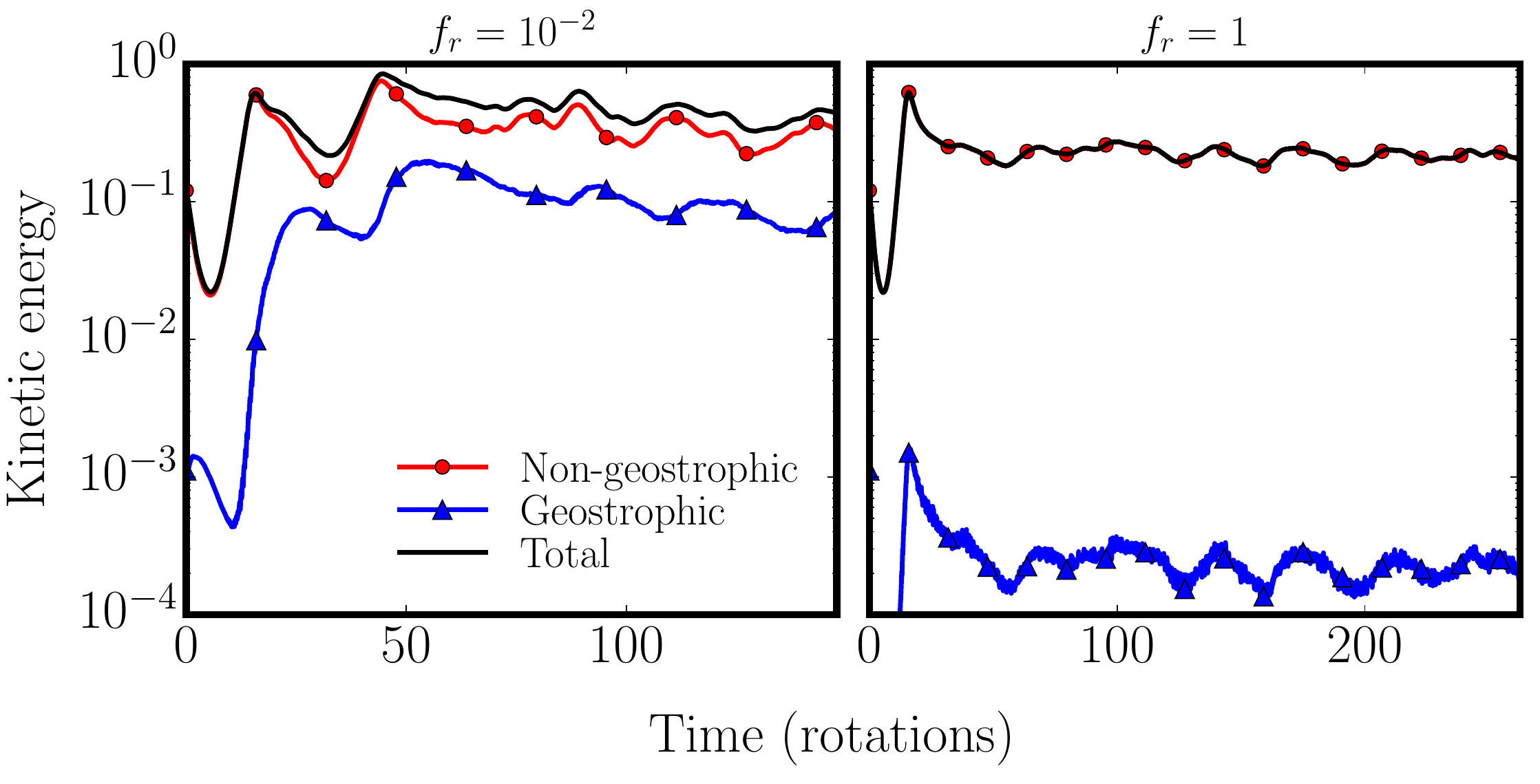}
    \includegraphics[width=0.49\linewidth]{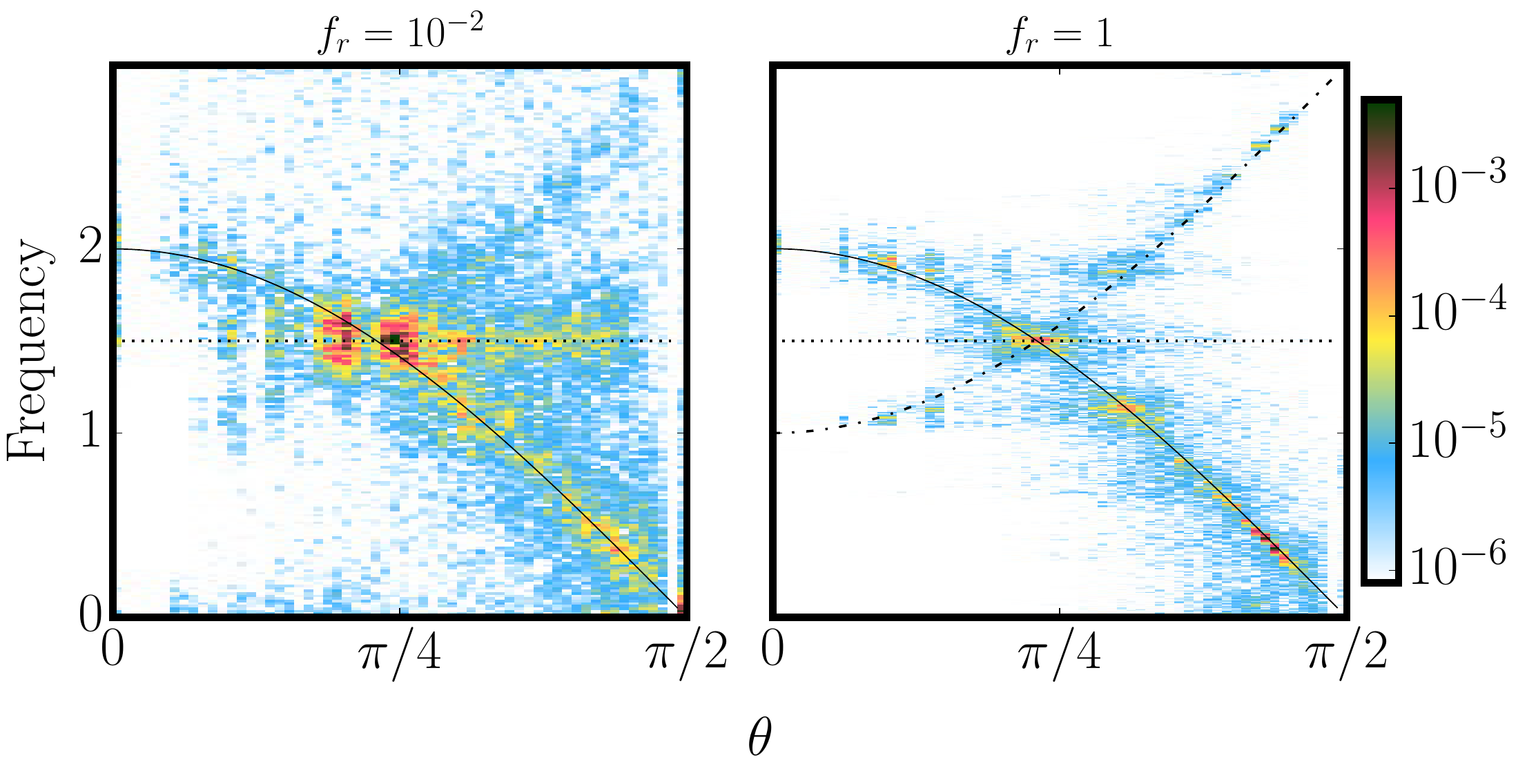}
    \vspace{-5mm}
    \caption{\textbf{Left:} contributions to the kinetic energy of the flow for two different values of the friction parameter with $E = 3 \times 10^{-6}$ and  $512^3$ resolution. \textbf{Right:} corresponding spectral energies processed over the saturated phase. 
    For $f_r = 1$, the dashed line gives the location of secondary non-resonant interactions of the wave with the base flow:  $-2\cos \theta+2 \gamma/\Omega$. The dotted horizontal lines locate the resonance frequency $\gamma/\Omega$.
    }
    \label{fig:fr12}
\end{figure*}

 In the following, we question the universality of these results.  
Our simulations are designed to study locally the properties of the flow in a wider container.
As the geostrophic modes are invariant along the rotation axis, they should connect with boundaries.
This interaction induces a secondary flow known as Ekman pumping, which acts as a bulk friction on these modes.
As is routinely done in quasi-geostrophic models of rapidly-rotating fluids \cite{schaeffer_quasigeostrophic_2005}, and in confined inertial wave turbulence \cite{scott2014}, an additionnal term is therefore added to the dynamics of the geostrophic part of the flow $ \bu_{\mathrm{G}} $, which reads (in appropiate units) as \cite{pedlosky_geophysical_1987}:
\textcolor{black}{%
\begin{equation}\label{friction}
    \mathcal{F}\left[ \bu_{\rm{G}} \right]  ~= ~ - \sqrt{E} ~ (L/h) ~\bu_{\mathrm{G}} ~=~ - \sqrt{E} f_r \bu_{\mathrm{G}} ~.
\end{equation}
}
Here $h$ is the height of the container along the rotation axis.
In Fourier space, (\ref{friction}) is tantamount to adding a term $- f_r E^{1/2} \hat{\bu}(\bk)$, for modes with $k_z = 0$.
Several additionnal reasons exist for controlling the growth of geostrophic modes. 
First, columnar vortices are typically not observed in global simulations. Instead, geostrophic modes emerge as steady axisymmetric flows, which do not necessarily inhibit the instability mechanism \cite{favier_generation_2015} (but see [20])
Moreover, the inverse cascade shown in Figure \ref{raw_results} inevitably leads to vortices of extent similar to the size of the box, which is unphysical.  
The same issue arises in two-dimensional turbulence, where it is solved adding large-scale friction \cite{boffetta_two-dimensional_2012}. 
Lastly, additionnal physics such as imposing a background magnetic field \cite{barker_mag_2014,guervilly2015}, can also be responsible for precluding the emergence of large scale geostrophic flows. 
The friction coefficient $f_r$, which stands for the ratio $L/h$, is left here as a parameter to control the relative importance of the geostrophic modes in the saturated flow.
We expect other mechanisms of specific dissipation (see \textit{e.g.} magnetic \cite{guervilly2015,barker_mag_2014}, quadratic, scale-dependant) to lead to similar results. 

The simulations hereafter are carried out  with $f_r = 10^{-2}$ and $f_r =1$. We now set the Ekman number $E$ to $3 \times 10^{-6}$ and the resolution to $512^3$.
These extreme parameters could not be reached previously since the transition from phase b. to c. (see Figure \ref{raw_results}) was then too difficult to resolve.
For comparison, the simulations with friction corresponding to Figure \ref{raw_results} can be found in Supplementary Materials.

\begin{figure*}
    \centering
    \includegraphics[width = 0.28\linewidth]{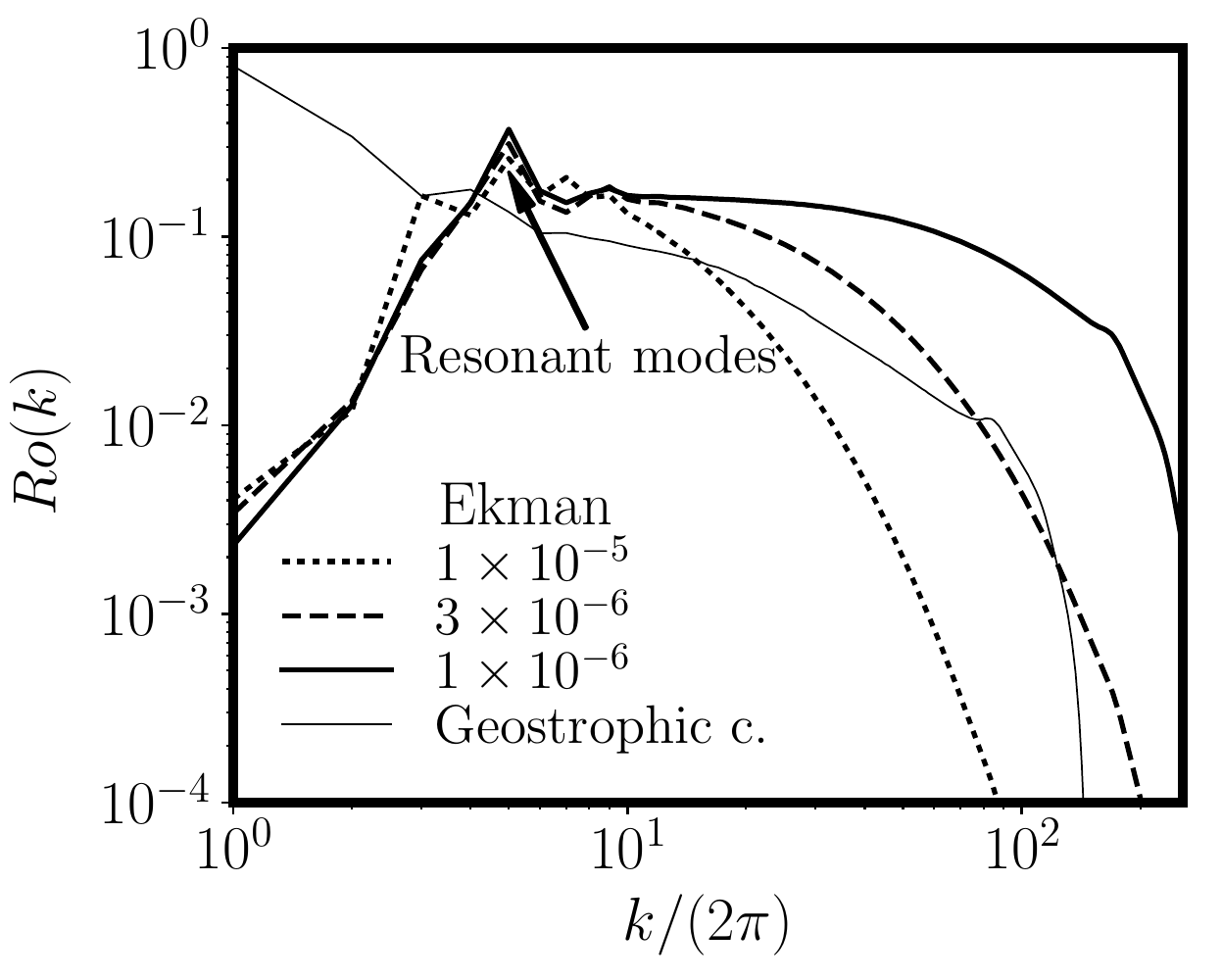}
    \includegraphics[width = 0.27\linewidth]{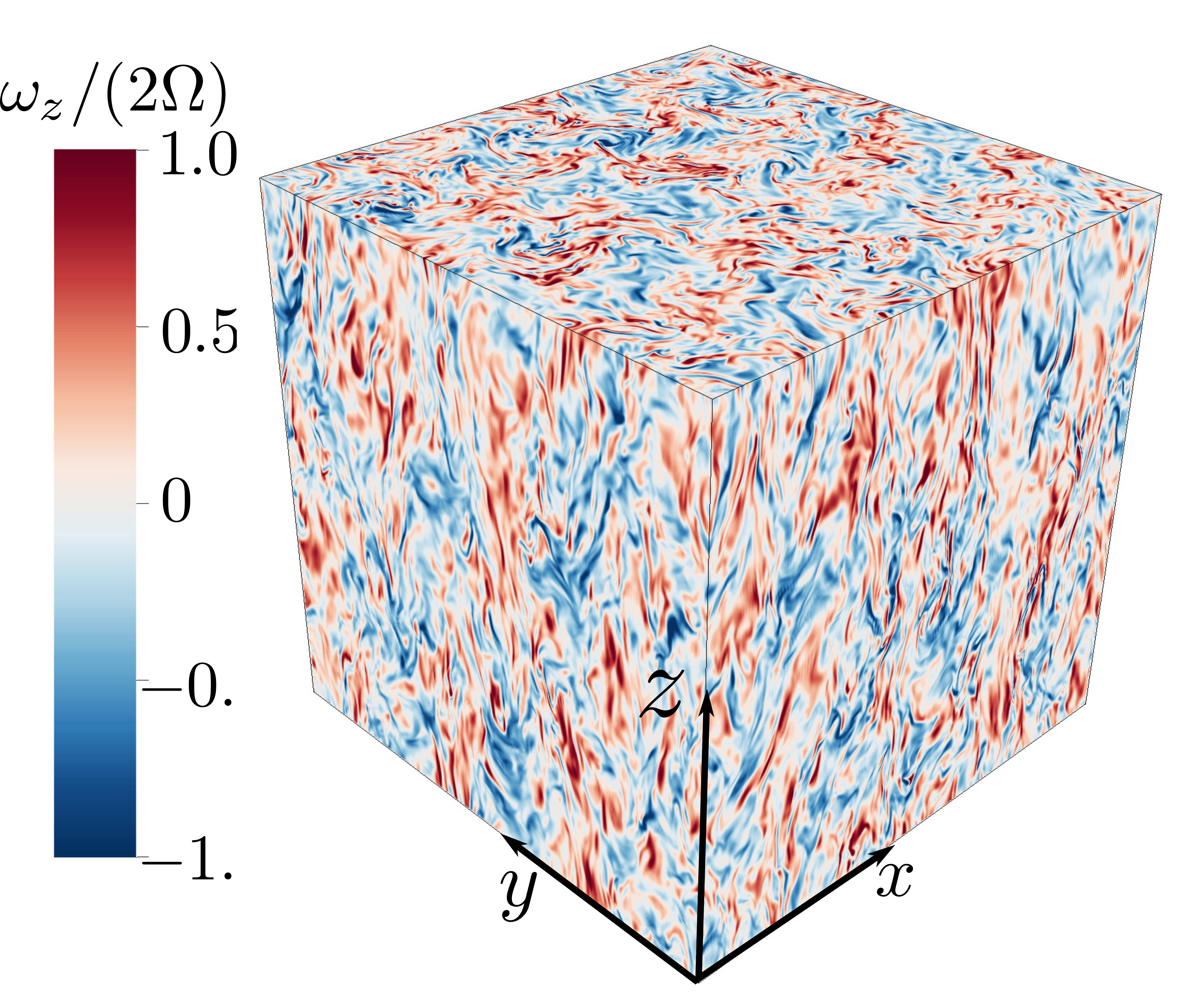}
    \includegraphics[width = 0.25\linewidth]{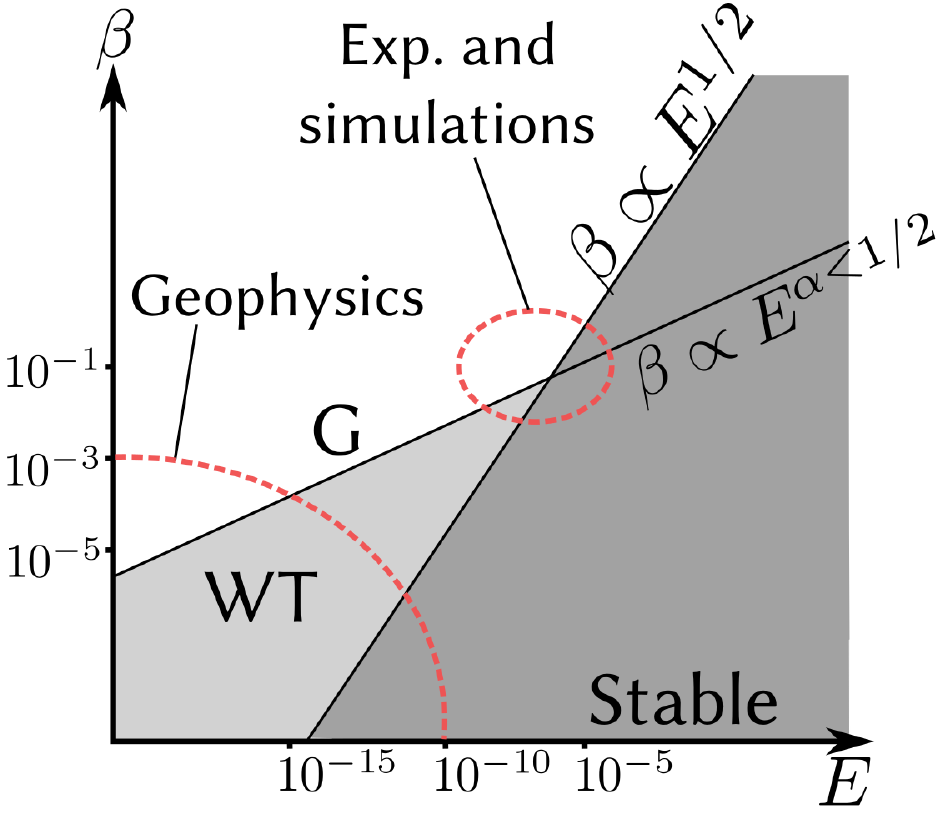}
    \caption{\textbf{Left:} scale-dependence of Rossby number for $\beta = 5 \times 10^{-2}$ and $f_r=1$ for $E= 10^{-5},\,3\times 10^{-6},\,10^{-6}$; results from the geostrophic saturation (Figure \ref{raw_results}c) are shown for comparison. 
    \textbf{Center:}  vertical vorticity in the $E = 10^{-6}$ case.
    \textbf{Right:} schematic diagram illustrating where the geostrophic and wave turbulence regimes can be expected depending on $\beta$ and $E$. WT and G stand for wave turbulence and geostrophic types of saturation, respectively.
    }

    \label{fig:spatial_spectra}
\end{figure*}

As displayed in Figure \ref{fig:fr12}, both relatively strong ($f_r = 1$) and weak ($f_r = 10^{-2}$) selective damping of the geostrophic modes leads to significant reduction of the geostrophic energy, but it does not affect the total energy, which remains similar to what was obtained previously in the early saturation phase (denoted as b. in Figure \ref{raw_results}). 
Moreover, the energy is maintained throughout the simulation.
The kinetic energy of the flow is now sharply located around the dispersion relation, 
not only at the resonance frequency, but at many frequencies consistent with the dispersion relation of inertial waves.
Other mirroring locations of the energy can be noticed (see dashed line in Figure \ref{fig:fr12}); they are understood as non-resonant interactions between the waves and the base flow with frequencies $ -2\cos \theta + 2 \gamma/\Omega$.
Energy focusing along the dispersion relation was observed in \cite{yarom_experimental_2014} but in their case energy was injected randomly into the system.
In our case, all the waves excited at frequencies different from the resonant frequency must be produced by nonlinear resonant interactions, hence energy focusing around the dispersion relation provides strong evidence to support the existence of inertial wave turbulence driven by elliptical instability.
We further investigate the creation of small scales when the geostrophic modes are sub-dominant by plotting the Rossby number as a function of the wavenumber $k$ 
computed as $Ro(k) = k E(k)^{1/2} / \Omega $, $E(k)$ being the isotropic energy spectrum. 
This is displayed in Figure \ref{fig:spatial_spectra} and suggests that for asympotically low Ekman numbers, $Ro(k)$ is constant and less than unity beyond the resonant scale.
Hence, non-linear interactions are weak and rotation affects all scales; this result is completely different from the Zeman phenomenology of forced rotating turbulence \cite{godeferd_structure_2015}, according to which isotropic Kolmogorov-like turbulence should be recovered at small scales.  

To conclude, our results not only clarify the various saturation regimes of the elliptical instability observed in experiments and numerical simulations, but also provide a new approach to disentangle wave and vortices in any situation involving rotating turbulence.
Depending on the relative importance of the geostrophic flows, one can observe intermittent behaviour, 
or quasi-stationary states close to inertial wave turbulence, provided that the input Rossby number (or equivalently $\beta$) is low enough.
As shown here, the fundamental quantity to be considered is therefore the ratio between geostrophic flows and 3D modes, which depends on the specifics of the considered system, and has been mostly neglected when comparing different rotating turbulence configurations. 
In previous experiments and simulations in spherical or ellipsoidal geometries, the geostrophic modes resulting from the non-linear interactions of inertial waves manifest themselves as zonal flows, \textit{i.e.} mean steady axisymmetric flows pervading the fluid interior \cite{noir_experimental_2012,morize_experimental_2010,grannan_submitted,grannan_experimental_2014,favier_generation_2015} (see \cite{lin_precession-driven_2016} however).
Their amplitudes tend to be proportional to $\beta^2E^{-\alpha}$ with $\alpha$ ranging from $0$ to~$2$ \cite{morize_experimental_2010,sauret_tide-driven_2014} depending on the excitation frequency. 
Since the amplitude of the RMS velocity scales like $\beta$ \cite{barker_non-linear_2013,grannan_submitted}, the ratio of the geostrophic zonal flows to the 3D modes is therefore proportional to $\beta E^{-\alpha}$. 
As the elliptical instability
grows provided $\beta > E^{1/2}$ \cite{le_bars_tidal_2010}, we conclude that for $\alpha<1/2$ there exists a regime where inertial wave turbulence is expected at low $\beta$ and $E$ --- see right pannel of Figure \ref{fig:spatial_spectra}.
Inertial wave turbulence can therefore be expected in planetary cores and more generally in rotating turbulent flows. 
This new turbulence regime has to be studied in detail in order to understand both dynamos \cite{moffatt_1970,bardsley_davidson_2016} and dissipation driven by the elliptical instability.
In the general framework of rotating fluid turbulence, our study indicates that mechanisms controlling the growth of geostrophic modes completely determine the nature of turbulence between quasi two-dimensional and wave turbulence.

\begin{acknowledgments} We acknowledge support from the European Research Council (ERC) under the European Union's Horizon 2020 research and innovation program (grant agreement No. 681835-FLUDYCO-ERC-2015-CoG). We also acknowledge support from IDRIS (Institut du D\'eveloppement et des Ressources en Informatique Scientifique) for computational time on Turing (Projects No. 100508 and 100614) and from the HPC resources of Aix-Marseille Universit\'e (Projects No.15b011 and 16b020) financed by the project Equip@Meso (No. ANR-10-EQPX-29-01) of the program Investissements d'Avenir supervised by the Agence Nationale pour la Recherche. AJB is supported by a Leverhulme Trust Early Career Fellowship.
\end{acknowledgments}

%

\end{document}